\documentclass[
10pt,
aps,
prl,
superscriptaddress,
secnumarabic,
showpacs,
intlimits,
amsmath,
amssymb,
floats,
floatfix,
twocolumn
]{revtex4-1}

\usepackage{bm}
\usepackage[final]{graphicx}
\usepackage{epsfig}
\usepackage[thinspace]{SIunits}
\usepackage{amsmath}
\usepackage{amssymb}
\usepackage{color}
\usepackage[colorlinks=true,citecolor=blue,linkcolor=blue,urlcolor=blue,pdfstartview=FitH,pdfauthor=Kretzschmar]{hyperref}

\begin{document}

\title{Raman-Scattering Detection of Nearly Degenerate $s$-Wave and $d$-Wave Pairing Channels in Iron-Based Ba$_{0.6}$K$_{0.4}$Fe$_2$As$_2$ and Rb$_{0.8}$Fe$_{1.6}$Se$_2$ Superconductors}
\date{\today}
\author{F. Kretzschmar}
\affiliation{Walther Meissner Institute, Bavarian Academy of Sciences and Humanities, 85748 Garching, Germany}
\author{B. Muschler}
\affiliation{Walther Meissner Institute, Bavarian Academy of Sciences and Humanities, 85748 Garching, Germany}
\author{T. B\"ohm}
\affiliation{Walther Meissner Institute, Bavarian Academy of Sciences and Humanities, 85748 Garching, Germany}
\author{A. Baum}
\affiliation{Walther Meissner Institute, Bavarian Academy of Sciences and Humanities, 85748 Garching, Germany}
\author{R. Hackl}
\affiliation{Walther Meissner Institute, Bavarian Academy of Sciences and Humanities, 85748 Garching, Germany}
%\author{T.P. Devereaux}
%\affiliation{Stanford Institute for Materials and Energy
%Sciences, SLAC National Accelerator Laboratory,
%2575 Sand Hill Road, Menlo Park, CA 94025, USA}
%\affiliation{Geballe Laboratory for Advanced Materials \& Dept. of Applied Physics,
%Stanford University, CA 94305, USA}
\author{Hai-Hu Wen}
\affiliation{National Laboratory of Solid State Microstructures and Department of Physics,
Nanjing University, Nanjing 210093, China}
\author{V. Tsurkan}
\affiliation{Experimantal Physics 5, Center for Electronic Correlations and Magnetism,
Institute of Physics, University of Augsburg, 86159 Augsburg, Germany}
\affiliation{Institute of Applied Physics, Academy of Sciences of Moldova, MD-2028,
Chisinau, Republic of Moldova}
\author{J. Deisenhofer}
\affiliation{Experimantal Physics 5, Center for Electronic Correlations and Magnetism,
Institute of Physics, University of Augsburg, 86159 Augsburg, Germany}
\author{A. Loidl}
\affiliation{Experimantal Physics 5, Center for Electronic Correlations and Magnetism,
Institute of Physics, University of Augsburg, 86159 Augsburg, Germany}
%\email{hackl@wmi.badw.de}

%%%%%%%%%%%%%%%%%%%%%%%%%%%%%%%%%%%%%%%%%%%%%%%%%%%%%%%%%%%%%%%%%%%%%%%%%%%%%%%%%%%%%%%%%
\begin{abstract}
We show that electronic Raman scattering affords a window into the essential properties of the pairing potential $V_{\textbf{k},\textbf{k}^{\prime}}$ of iron-based superconductors. In ${\rm Ba_{0.6}K_{0.4}Fe_2As_2}$ we observe band dependent energy gaps along with excitonic Bardasis-Schrieffer modes characterizing, respectively, the dominant and subdominant pairing channel. The $d_{x^2-y^2}$ symmetry of all excitons allows us to identify the subdominant channel to originate from the interaction between the electron bands. Consequently, the dominant channel driving superconductivity results from the interaction between the electron and hole bands and has the full lattice symmetry. The results in ${\rm Rb_{0.8}Fe_{1.6}Se_2}$ along with earlier ones in ${\rm Ba(Fe_{0.939}Co_{0.061})_2As_2}$ highlight the influence of the Fermi surface topology on the pairing interactions.
\end{abstract}
%%%%%%%%%%%%%%%%%%%%%%%%%%%%%%%%%%%%%%%%%%%%%%%%%%%%%%%%%%%%%%%%%%%%%%%%%%%%%%%%%%%%%%%%%
\maketitle
%%%%%%%%%%%%%%%%%%%%%%%%%%%%%%%%%%%%%%%%%%%%%%%%%%%%%%%%%%%%%%%%%%%%%%%%%%%%%%%%%%%%%%%%%

\begin{figure}[hbtp]
  \centering
  \includegraphics[width=8.0cm]{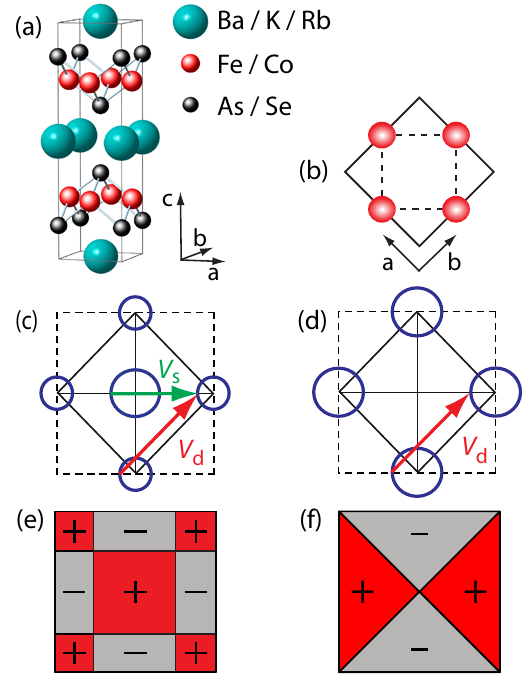}
  %\vspace{-0.8cm}
  \caption[]{(Color online)
Crystal and reciprocal lattices of iron-based superconductors. (a) FeAs/Se layers (small spheres) and (earth) alkali metals (big spheres). (b) 1\,Fe (dashes) and 2\,Fe (full line) unit cells. (c), (d) First Brillouin zones (BZ, dashes for the 1\,Fe cell) with schematic Fermi surfaces of (c), ${\rm Ba_{0.6}K_{0.4}Fe_2As_2}$ \cite{Ding:2008, Evtushinsky:2009a} and (d), ${\rm Rb_{0.8}Fe_{1.6}Se_2}$ \cite{Zhang:2011a,Qian:2011}. Also indicated are the two dominant interaction potentials $V_s$ and $V_d$. (e) For $V_s > V_d$ the pairing state is predicted to have $s_{\pm}$ symmetry for which the phase of the gap on the hole and electron bands differs by $\pi$ \cite{Mazin:2008a,Thomale:2011}. (f) For $V_d > V_s$ a state with $d_{x^2-y^2}$ symmetry is favored \cite{Lee:2009,Scalapino:2009}.
}
  \label{fig:structure}
\end{figure}

Cooper pairing in superconductors is driven by the interaction potential $V_{\textbf{k},\textbf{k}^{\prime}}$ between two electrons. In conventional superconductors with an isotropic gap $\Delta$ prominent structures appear in many spectroscopies at $\hbar\omega_{\textbf{q}}+\Delta$, and $V_{\textbf{k},\textbf{k}^{\prime}}$ can be derived by and large from the spectrum of interactions $\hbar\omega_{\textbf{q}}$ \cite{McMillan:1965}. This access is hampered in systems with the gap $\Delta_{\textbf{k}}$ varying strongly with the electronic momentum $\hbar\textbf{k}$. The iron-based superconductors \cite{Kamihara:2008,Rotter:2008a}, as shown in Fig.~\ref{fig:structure}, open up new vistas. Since the hole- and electron-like Fermi surfaces can be tuned by substitution [Fig.~\ref{fig:structure}(c) and (d)] they can be considered model systems for studying the pairing interaction in anisotropic multi-band systems \cite{Hirschfeld:2011,Lee:2009,Thomale:2011}. Repulsive spin \cite{Mazin:2008a} and attractive orbital \cite{Kontani:2010} fluctuations were suggested to provide appreciable interaction potentials $V_{\textbf{k},\textbf{k}^{\prime}}$. The resulting ground states may preserve [Fig.~\ref{fig:structure}(e)] or break the full lattice symmetry [Fig.~\ref{fig:structure}(f)]. In the spin channel, the interactions between either the central hole-like and the peripheral electron-like Fermi surfaces $V_s$ \cite{Mazin:2008a} or the electron bands alone $V_d$ are nearly degenerate \cite{Lee:2009,Scalapino:2009} [Fig.~\ref{fig:structure}(c) and (d)] and entail a sign change of the energy gap $\Delta_{\textbf{k}}$ [Fig.~\ref{fig:structure}(e) and (f)].

Raman scattering offers an opportunity to scrutinize competing superconducting instabilities and derive essential properties of $V_{\textbf{k},\textbf{k}^{\prime}}$. The electronic response provides direct access to the energy gap and its momentum dependence \cite{Devereaux:1994,Devereaux:2007} reflecting the dominant channel responsible for Cooper pairing. In addition, residual interactions resulting from anisotropies of the pairing potential $V_{\textbf{k},\textbf{k}^{\prime}}$ may lead to sharp in-gap modes due to the formation of bound states of the two electrons of a broken Cooper pair \cite{Bardasis:1961,Klein:1984,Monien:1990,Scalapino:2009,Chubukov:2009a} similar to electron-hole excitons in semiconductors. The energy and the symmetry properties of these ``Cooperons'' provide insight into the momentum dependence of $V_{\textbf{k},\textbf{k}^{\prime}}$ or decompositions thereof in terms of orthonormal functions $\phi_i$ such as $V_{\textbf{k},\textbf{k}^{\prime}} = \phi^2_s V_s + \phi^2_d V_d + \dots$ and, consequently, the type of interaction.

%%%%%%%%%%%%%%%%%%%%%%%%%%%%%%%%%%%%%%%%%%%%%%%%%%%%%%%%%%%%%%%%%%%%%%%%%%%%%%%%%%%%%%%%%
%%%%%%%%%%%%%%%%%%%%%%%%%%%%%%%%%%%%%%%%%%%%%%%%%%%%%%%%%%%%%%%%%%%%%%%%%%%%%%%%%%%%%%%%%

In this paper, we present light scattering spectra as a function of the Fermi surface topology in order to gain insight into the pairing interaction. The results on the electron-doped iron-chalcogenide ${\rm Rb_{0.8}Fe_{1.6}Se_2}$ and the hole-doped iron-pnictide ${\rm Ba_{0.6}K_{0.4}Fe_2As_2}$ are analyzed along with the data of ${\rm Ba(Fe_{0.939}Co_{0.061})_2As_2}$ studied earlier \cite{Muschler:2009}.

The ${\rm Ba_{0.6}K_{0.4}Fe_2As_2}$ sample has a transition temperature $T_c$ of 39\,K. Good crystallinity and low defect concentration were shown by x-ray diffraction and specific heat measurements \cite{Shen:2011}. ${\rm Rb_{0.8}Fe_{1.6}Se_2}$ has a sharp transition at $T_c=32$\,K \cite{Tsurkan:2011}. The recently discovered layered structure \cite{Ksenofontov:2011} has little influence on the light scattering experiments since only the metallic (superconducting) part contributes to the particle-hole continuum.

\begin{figure}[hbtp]
  \centering
  \includegraphics[width=8.0cm]{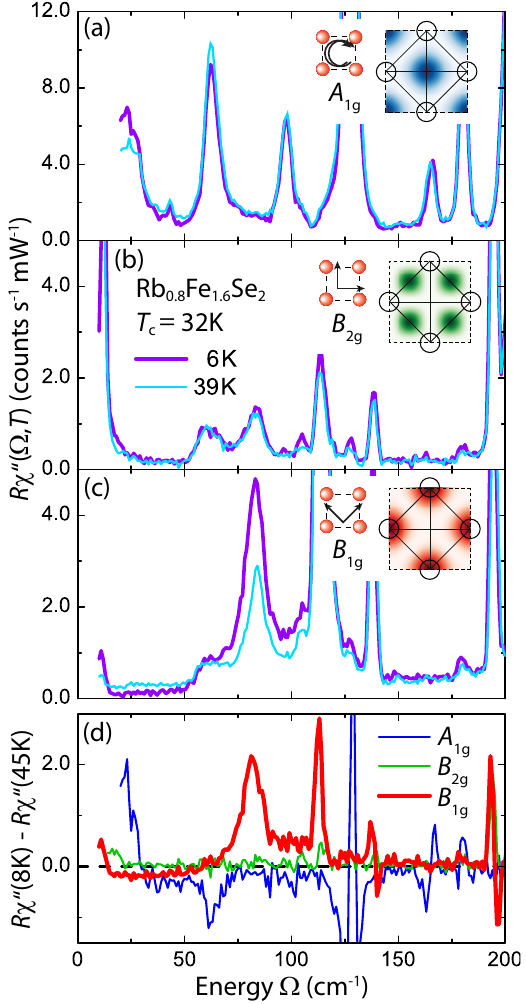}
  %\vspace{-0.8cm}
  \caption[]{(Color online)
Normal and superconducting Raman spectra of ${\rm Rb_{0.8}Fe_{1.6}Se_2}$ at temperatures as indicated. The insets show the correspondence between light polarizations and sensitivities in momentum space for the 1\,Fe unit cell. (a)-(c) Spectra in $A_{1g}$,  $B_{2g}$, and $B_{1g}$ symmetry. (d) The difference spectra highlight the absence of pair-breaking in $B_{2g}$ and most likely also in $A_{1g}$ symmetry. Only the $B_{1g}$ spectra show the features typical for a superconductor.
  }
  \label{fig:RFS_data}
\end{figure}

\begin{figure}[hbtp]
  \centering
  \includegraphics[width=8.0cm]{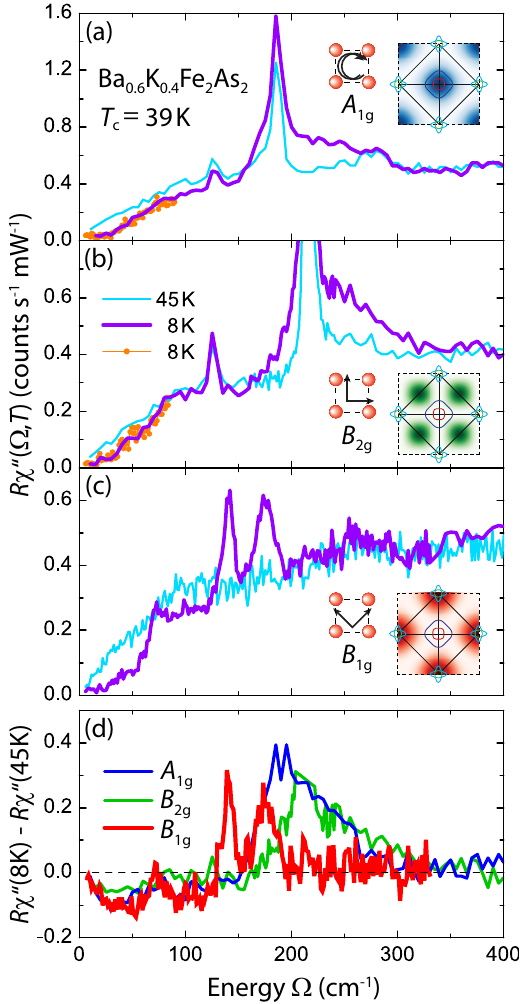}
  %\vspace{-0.8cm}
  \caption[]{(Color online)
Normal and superconducting Raman spectra of ${\rm Ba_{0.6}K_{0.4}Fe_2As_2}$ at temperatures as indicated. The spectra plotted with full lines in panel (a) and (b) are measured with a resolution of 6.5\,cm$^{-1}$ whereas the spectra in panel (c) and the spectra displayed with orange points are measured with a resolution of 4.7\,cm$^{-1}$. The insets show the correspondence between light polarizations and sensitivities in momentum space for the 1\,Fe unit cell. (a)-(c)  Spectra in $A_{1g}$, $B_{2g}$, and $B_{1g}$ symmetries. (d) Difference between superconducting and normal-state spectra. In $B_{2g}$ symmetry (green), the phonon at 215\,cm$^{-1}$ is truncated for having a higher intensity below $T_c$.
  }
  \label{fig:BKFA_data}
\end{figure}

The experiments were performed with standard light scattering equipment \cite{Devereaux:2007} using a solid state laser emitting at 532\,nm. The ${\rm Rb_{0.8}Fe_{1.6}Se_2}$ crystal was cleaved \textit{in situ} at low temperature. The figures display the Raman susceptibilities $R\chi^{\prime\prime}_{\gamma,\gamma}(\Omega,T)=S(\Omega,T)\{1+n(T,\Omega)\}^{-1}$ where $R$ is an experimental constant, $S$ is the van Hove function being proportional to the rate of scattered photons, and $n$ is the Bose-Einstein distribution.  The polarizations of the incoming and scattered photons are given with respect to the 1\,Fe unit cell [Fig.~\ref{fig:structure}\,(b)] relevant for the electronic properties [for details see Supplemental Material (SM)]. The related excitation symmetries translate into sensitivities in momentum space for electron-hole excitations \cite{Devereaux:2007} as shown in the insets of Figs.~\ref{fig:RFS_data} and \ref{fig:BKFA_data}. The symmetry properties of the collective modes reflect those of the subdominant channels in the potential $V_{\textbf{k},\textbf{k}^{\prime}}$ which do not support pairing in the ground state \cite{Monien:1990}.

The symmetry-dependent Raman response of ${\rm Rb_{0.8}Fe_{1.6}Se_2}$ is shown in Fig.~\ref{fig:RFS_data}. Due to surface issues (see SM) there is a relatively strong increase towards the laser line. In the $A_{1g}$ and $B_{2g}$ spectra, the relative difference between the normal and the superconducting state is weak and absent, respectively, since the band structure of ${\rm Rb_{0.8}Fe_{1.6}Se_2}$ \cite{Zhang:2011a,Qian:2011} does not have Fermi surface crossings close to the sensitivity maxima of the related form factors (see insets of Fig.~\ref{fig:RFS_data}).
In $B_{1g}$ symmetry, the suppression of  the low-temperature spectra due to the gap and the excess intensity at and above $2\Delta$ can be considered typical features of a superconductor \cite{Devereaux:2007}. The relative changes of below and above a threshold at approximately 60\,cm$^{-1}$ reach 80 and 30\%, respectively [Fig.~\ref{fig:RFS_data}(c)].  Below 60\,cm$^{-1}$, the intensity is only weakly energy dependent indicating a clean isotropic gap. The phonons at 80 and 115\,cm$^{-1}$, close to the gap edge, gain intensity below $T_c$ as expected for weak electron-phonon coupling.
As can be seen directly in the insets of Fig.~\ref{fig:RFS_data} an appreciable response is expected only in $B_{1g}$ symmetry if hole-like bands in the Brillouin zone center are absent. In the presence of hole bands, gap structures appear also in $A_{1g}$ symmetry \cite{Muschler:2009}. Hence, the selection rules based on symmetry arguments corresponding to the 1\,Fe unit cell \cite{Muschler:2009} are supported by the results in ${\rm Rb_{0.8}Fe_{1.6}Se_2}$ and are likely to be of general significance in the iron-based superconductors.

%%%%%%%%%%%%%%%%%%%%%%%%%%%%%%%%%%%%%%%%%%%%%%%%%%%%%%%%%%%%%%%%%%%%%%%%%%%%%%%%%%%%%%%%%
%%%%%%%%%%%%%%%%%%%%%%%%%%%%%%%%%%%%%%%%%%%%%%%%%%%%%%%%%%%%%%%%%%%%%%%%%%%%%%%%%%%%%%%%%

In Fig.~\ref{fig:BKFA_data}, we show Raman spectra of ${\rm Ba_{0.6}K_{0.4}Fe_2As_2}$. Here, we observe superconductivity-induced features in all symmetries. Below a symmetry independent threshold of approximately 25\,cm$^{-1}$ the response is very small and nearly energy independent. Although the intensity is not exactly zero it is safe to conclude that there is a full gap on all bands having a magnitude of at least 0.9\,$k_BT_c$. The excess intensity in the range $130 < \Omega < 300$\,cm$^{-1}$ originates from either pair-breaking [Fig.~\ref{fig:BKFA_data}(a) and (b)] or, as will be discussed in detail below, from collective excitations [Fig.~\ref{fig:BKFA_data}(c)]. The spectral features in $A_{1g}$ and $B_{2g}$  symmetry [Fig.~\ref{fig:BKFA_data}(a) and (b)] are broad and asymmetric whereas the peaks in $B_{1g}$ symmetry [Fig.~\ref{fig:BKFA_data}(c)] are rather sharp, even though not resolution limited, and symmetric. The important secondary structures of the spectra between the minimal and the maximal gaps are better resolved in the difference spectra in panel (d) of Fig.~\ref{fig:BKFA_data}. In this way, the contributions from phonons are by and large subtracted out. Only the Fe vibration at 215\,cm$^{-1}$ in the $B_{2g}$ spectrum ($B_{1g}$ phonon in the crystallographic cell) has an anomalous intensity for its proximity to the gap edge and is truncated. Both the $A_{1g}$ and $B_{2g}$ spectra have edge-like onsets above approximately 150\,cm$^{-1}$ before reaching distinct maxima at 190 and 210\,cm$^{-1}$, respectively, and decaying slowly.

\begin{figure}[hbtp]
  \centering
  \includegraphics[width=8.0cm]{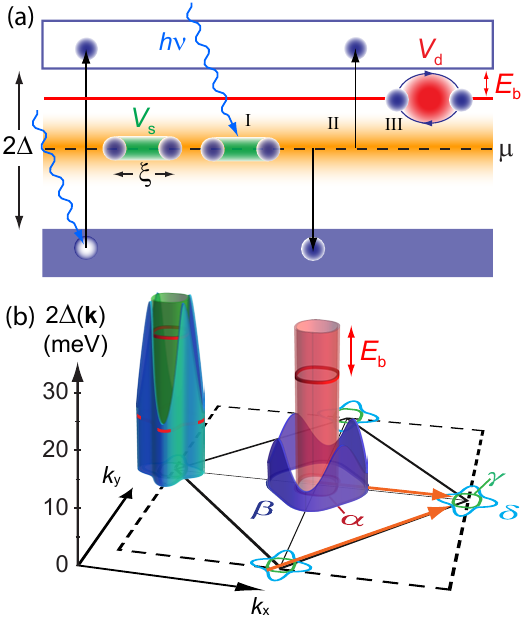}
  %\vspace{-0.8cm}
  \caption[]{(Color online)
Energy gaps and excitons in superconductors. (a) Mechanism for Bardasis-Schrieffer excitonic modes. (b) Most probable anisotropy of the energy gap in ${\rm Ba_{0.6}K_{0.4}Fe_2As_2}$. In the SM arguments are provided how the gap distribution shown here is derived from Fig.~\ref{fig:BKFA_data}. The energy of the bound state $E_b$ (red line) is largest on the $\alpha$ band.
  }
  \label{fig:BKFA_exciton}
\end{figure}

In ${\rm Ba_{0.6}K_{0.4}Fe_2As_2}$, in contrast to ${\rm Rb_{0.8}Fe_{1.6}Se_2}$ and ${\rm Ba(Fe_{0.939}Co_{0.061})_2As_2}$ (Refs.~\cite{Muschler:2009,Chauviere:2010}), gap features are observed also in $B_{2g}$ symmetry. At first glance, this appears to be at odds with the selection rules. However, the outer hole ($\beta$) band has a large Fermi momentum $k_F(\beta)$ at which the $B_{2g}$ vertex reaches already 20\% of its maximum. In addition, the $B_{2g}$ vertex is enhanced by a factor of approximately 2-5  at the hybridization points of the electron-like $\gamma$ and $\delta$ bands \cite{Mazin:2010a}. Therefore, one can see maxima of both the small gap of the $\beta$ band and the large gap on the $\gamma/\delta$ bands at 80 and 210\,cm$^{-1}$, respectively. Apparently, the gap is large at the hybridization points in ${\rm Ba_{0.6}K_{0.4}Fe_2As_2}$ as opposed to ${\rm Ba(Fe_{0.939}Co_{0.061})_2As_2}$ \cite{Muschler:2009,Mazin:2010a}. The gap on the outer electron ($\delta$) band cannot be large everywhere. Otherwise the $A_{1g}$ and $B_{2g}$ spectra would not have distinctly different maxima since the $A_{1g}$  vertex is finite everywhere on the $\delta$ band except for the degeneracy points. Hence, the kink at 80\,cm$^{-1}$ in $A_{1g}$ symmetry indicates that the gap on the $\delta$ band varies between 80 and 210\,cm$^{-1}$. Finally, the minimal gap of 25\,cm$^{-1}$ is expected to reside in the $\beta$ band along the principle axes since it is observed in all symmetries. Upon including the $B_{1g}$ collective modes at 140 and 175\,cm$^{-1}$, which exist only inside clean gaps \cite{Scalapino:2009}, it follows directly that the gaps on the $\alpha$ and the $\gamma$ bands are fairly isotropic. The $B_{1g}$ maximum at 73\,cm$^{-1}$ is the collective mode pulled down from the minimal gap of the $\delta$ band \cite{Scalapino:2009}. Hence, the comparison of the spectra in $A_{1g}$ and $B_{2g}$ symmetry and the positions and shapes of the $B_{1g}$ collective excitations allow us to derive a consistent proposal for how the gaps most likely vary on the individual bands as shown in Fig.~\ref{fig:BKFA_exciton}\,(b). Further details of the derivation are described in the SM.

%%%%%%%%%%%%%%%%%%%%%%%%%%%%%%%%%%%%%%%%%%%%%%%%%%%%%%%%%%%%%%%%%%%%%%%%%%%%%%%%%%%%%%%%%
%%%%%%%%%%%%%%%%%%%%%%%%%%%%%%%%%%%%%%%%%%%%%%%%%%%%%%%%%%%%%%%%%%%%%%%%%%%%%%%%%%%%%%%%%

At this point, a more general picture emerges from the synopsis of the results in ${\rm Ba_{0.6}K_{0.4}Fe_2As_2}$, ${\rm Ba(Fe_{0.939}Co_{0.061})_2As_2}$, and  ${\rm Rb_{0.8}Fe_{1.6}Se_2}$ with and without central hole bands. In the two extreme cases studied here, we observe more isotropic gaps whereas a strong anisotropy was found in ${\rm Ba(Fe_{0.939}Co_{0.061})_2As_2}$ \cite{Hirschfeld:2011,Muschler:2009} indicating a dramatic change of the pairing potential as predicted in various theoretical studies \cite{Platt:2012, Kuroki:2009, Ikeda:2010, Maiti:2011a, Das:2012, Fernandes:2012a}.

We now provide evidence for a competition between the two possible pairing states resulting from the exchange of spin fluctuations with $V_s$ and $V_d$ [see Fig.~\ref{fig:structure}(c)] winning in ${\rm Ba_{0.6}K_{0.4}Fe_2As_2}$ and  ${\rm Rb_{0.8}Fe_{1.6}Se_2}$, respectively. $V_s$ dominates but $V_d$ is appreciable in ${\rm Ba_{0.6}K_{0.4}Fe_2As_2}$  \cite{Thomale:2011a}. Then, one expects the system to condense into a ground state having the full symmetry of the lattice and to additionally develop $\delta$-like collective modes with $d_{x^2-y^2}$ orbital momentum bound by the residual interaction $V_d$ inside the gap as shown schematically in Fig.~\ref{fig:BKFA_exciton}(a). Photons ($h\nu$) scatter from both unpaired electrons in the Bogoliubov quasiparticle bands (left) and Cooper pairs at the chemical potential $\mu$ (right) \cite{Abrikosov:1961}. In either case an energy of at least $2\Delta$ must be invested, and an electron-hole pair and two unpaired electrons are created, respectively. The two electrons being separated by $2\Delta$ but remaining in a volume characterized by the coherence length $\xi_0$ ``sense'' now those parts of the interaction potential $V_{\textbf{k},\textbf{k}^{\prime}}$ which are orthogonal to the pairing channel and form a bound state of energy $E_b$ inside the gap \cite{Bardasis:1961,Klein:1984,Monien:1990,Devereaux:2007,Scalapino:2009,Chubukov:2009a}. The $\delta$-like modes appear below the gap edge at $\Omega_b=2\Delta-E_b$ with $E_b$ being the binding energy of the ``Cooperon'' [Fig.~\ref{fig:BKFA_exciton}(a)] \cite{Klein:1984,Monien:1990}. $E_b$ encodes the coupling strength in the subdominant channel and $E_b/2\Delta\approx (V_d/V_s)^2$. The modes at 140, and 175\,cm$^{-1}$ correspond to the $A_{1g}$ and $B_{2g}$ gap structures at 190 and 210\,cm$^{-1}$ implying binding energies $E_b$ of 50, and 35\,cm$^{-1}$ [Fig.~\ref{fig:BKFA_exciton}(b)], respectively, or 25\% of $2\Delta$ and indicate that $V_d$ is smaller but of the same order of magnitude as $V_s$ \cite{Monien:1990}. The bound state at 73\,cm$^{-1}$ corresponds to the minimum at 80\,cm$^{-1}$ of the strongly momentum dependent gap on the $\delta$ band as predicted by Scalapino and Devereaux \cite{Scalapino:2009}. We note that the mode at 140\,cm$^{-1}$ corresponding to the gap on the $\alpha$ band is unlikely to result from the interband coupling discussed in Ref.~\cite{Scalapino:2009}. Rather, it may originate in intraband coupling due to the $d_{x^2-y^2}$ component of $V_{\textbf{k},\textbf{k}^{\prime}}$, similarly as derived for one band superconductors, and is expected to appear in $B_{1g}$ symmetry as the other collective modes \cite{Klein:1984,Monien:1990}. This scenario explains in a natural way the two well-defined symmetric peaks at 140 and 175\,cm$^{-1}$ and the weaker one at 73\,cm$^{-1}$ appearing in (the proper) $B_{1g}$ symmetry [Fig.~\ref{fig:BKFA_data}(c) and (d)] and provides direct evidence of a strongly anisotropic pairing potential resulting from a superposition of $V_s$ and $V_d$.  Since the collective modes drain intensity from the gap features \cite{Monien:1990,Scalapino:2009} direct pair breaking peaks cannot be resolved in $B_{1g}$ symmetry. This effect provides the most compelling evidence for the proper interpretation of the in-gap modes.

Upon going from ${\rm Ba_{0.6}K_{0.4}Fe_2As_2}$ to ${\rm Ba(Fe_{0.939}Co_{0.061})_2As_2}$ $V_d$ increases becoming comparable to  $V_s$. At first glance the collective modes are expected to become stronger. However, more dramatic changes occur, and the entire pairing state becomes sufficiently anisotropic to drive the minimal gap almost the whole way down to zero at least on portions of the Fermi surface \cite{Hirschfeld:2011,Muschler:2009}. Now, the collective modes are heavily damped and almost undetectable within experimental uncertainties \cite{Scalapino:2009,Devereaux:1995a}.

%%%%%%%%%%%%%%%%%%%%%%%%%%%%%%%%%%%%%%%%%%%%%%%%%%%%%%%%%%%%%%%%%%%%%%%%%%%%%%%%%%%%%%%%%
%%%%%%%%%%%%%%%%%%%%%%%%%%%%%%%%%%%%%%%%%%%%%%%%%%%%%%%%%%%%%%%%%%%%%%%%%%%%%%%%%%%%%%%%%

If spin fluctuations dominate the coupling \cite{Mazin:2008a} $V_s$ is vanishingly small in ${\rm Rb_{0.8}Fe_{1.6}Se_2}$ because the central Fermi surface is missing. With $V_d$ surviving alone, the resulting pairing is $d_{x^2-y^2}$ without nodes since the sign change of the gap occurs on the Brillouin zone diagonals far away from the Fermi surface [Fig.~\ref{fig:structure}(f)]. However, after down-folding the 1\,Fe Brillouin zone (see Fig.~\ref{fig:structure}) the electron bands accommodating gaps having opposite sign are expected to hybridize. Without mixing of the Cooper pairs on the different bands the gaps have to change sign on the hybridization lines \cite{Mazin:2011a}. The resulting nodes on the Fermi surface are incompatible with the observation of a clean gap [Fig.~\ref{fig:RFS_data}(c)]. If, however, the Cooper pairs can couple across the bands \cite{Khodas:2012} nodeless states can occur again. Our observation of a clean gap in ${\rm Rb_{0.8}Fe_{1.6}Se_2}$ favors strong hybridization with almost circular concentric Fermi surfaces \cite{Khodas:2012}. As an immediate consequence there is no enhancement of the $B_{2g}$ Raman vertices in ${\rm Rb_{0.8}Fe_{1.6}Se_2}$ (as opposed to ${\rm Ba_{0.6}K_{0.4}Fe_2As_2}$). This explains both the absence of $B_{2g}$ gap features and the clean gap in ${\rm Rb_{0.8}Fe_{1.6}Se_2}$.

In conclusion, the analysis of the Raman results on the gap and of the excitonic in-gap modes of superconducting ${\rm Ba_{0.6}K_{0.4}Fe_2As_2}$ provides access to the anisotropy of the pairing potential $V_{\textbf{k},\textbf{k}^{\prime}}$. The dependence of the components $V_s$ and $V_d$ of $V_{\textbf{k},\textbf{k}^{\prime}}$ on the Fermi surface topology makes a strong case for pairing mediated by spin fluctuations. In this way, Raman scattering may well be useful in other circumstances as suggested recently by Barlas and Varma \cite{Barlas:2012} and supplement other methods.

\section*{Acknowledgements}
We gratefully acknowledge discussions with A. Chubukov, T.P. Devereaux, D. Einzel, W. Hanke, A.L. Kemper, I. Mazin, B. Moritz, E.A. Nowadnick, and D.J. Scalapino. R.H. is indebted to the Stanford Institute for Materials and Energy Sciences for the hospitality. We acknowledge support by the DFG under grant number Ha\,2071/7 in the Priority Program SPP\,1458 as well as via the Collaborative Research Center TRR\,80. The work in China is supported by the NSF of China, the Ministry of Science and Technology of China (973 projects: 2011CBA00102 and 2012CB821403).

%\bibliography{literature_rudi}

\begin{thebibliography}{37}%
\makeatletter
\providecommand \@ifxundefined [1]{%
 \@ifx{#1\undefined}
}%
\providecommand \@ifnum [1]{%
 \ifnum #1\expandafter \@firstoftwo
 \else \expandafter \@secondoftwo
 \fi
}%
\providecommand \@ifx [1]{%
 \ifx #1\expandafter \@firstoftwo
 \else \expandafter \@secondoftwo
 \fi
}%
\providecommand \natexlab [1]{#1}%
\providecommand \enquote  [1]{``#1''}%
\providecommand \bibnamefont  [1]{#1}%
\providecommand \bibfnamefont [1]{#1}%
\providecommand \citenamefont [1]{#1}%
\providecommand \href@noop [0]{\@secondoftwo}%
\providecommand \href [0]{\begingroup \@sanitize@url \@href}%
\providecommand \@href[1]{\@@startlink{#1}\@@href}%
\providecommand \@@href[1]{\endgroup#1\@@endlink}%
\providecommand \@sanitize@url [0]{\catcode `\\12\catcode `\$12\catcode
  `\&12\catcode `\#12\catcode `\^12\catcode `\_12\catcode `\%12\relax}%
\providecommand \@@startlink[1]{}%
\providecommand \@@endlink[0]{}%
\providecommand \url  [0]{\begingroup\@sanitize@url \@url }%
\providecommand \@url [1]{\endgroup\@href {#1}{\urlprefix }}%
\providecommand \urlprefix  [0]{URL }%
\providecommand \Eprint [0]{\href }%
\providecommand \doibase [0]{http://dx.doi.org/}%
\providecommand \selectlanguage [0]{\@gobble}%
\providecommand \bibinfo  [0]{\@secondoftwo}%
\providecommand \bibfield  [0]{\@secondoftwo}%
\providecommand \translation [1]{[#1]}%
\providecommand \BibitemOpen [0]{}%
\providecommand \bibitemStop [0]{}%
\providecommand \bibitemNoStop [0]{.\EOS\space}%
\providecommand \EOS [0]{\spacefactor3000\relax}%
\providecommand \BibitemShut  [1]{\csname bibitem#1\endcsname}%
\let\auto@bib@innerbib\@empty
%</preamble>
\bibitem [{\citenamefont {McMillan}\ and\ \citenamefont
  {Rowell}(1965)}]{McMillan:1965}%
  \BibitemOpen
  \bibfield  {author} {\bibinfo {author} {\bibfnamefont {W.~L.}\ \bibnamefont
  {McMillan}}\ and\ \bibinfo {author} {\bibfnamefont {J.~M.}\ \bibnamefont
  {Rowell}},\ }\href {\doibase 10.1103/PhysRevLett.14.108} {\bibfield
  {journal} {\bibinfo  {journal} {Phys. Rev. Lett.}\ }\textbf {\bibinfo
  {volume} {14}},\ \bibinfo {pages} {108} (\bibinfo {year} {1965})}\BibitemShut
  {NoStop}%
\bibitem [{\citenamefont {Kamihara}\ \emph {et~al.}(2008)\citenamefont
  {Kamihara}, \citenamefont {Watanabe}, \citenamefont {Hirano},\ and\
  \citenamefont {Hosono}}]{Kamihara:2008}%
  \BibitemOpen
  \bibfield  {author} {\bibinfo {author} {\bibfnamefont {Y.}~\bibnamefont
  {Kamihara}}, \bibinfo {author} {\bibfnamefont {T.}~\bibnamefont {Watanabe}},
  \bibinfo {author} {\bibfnamefont {M.}~\bibnamefont {Hirano}}, \ and\ \bibinfo
  {author} {\bibfnamefont {H.}~\bibnamefont {Hosono}},\ }\href {\doibase
  10.1021/ja800073m} {\bibfield  {journal} {\bibinfo  {journal} {J. Am. Chem.
  Soc.}\ }\textbf {\bibinfo {volume} {130}},\ \bibinfo {pages} {3296} (\bibinfo
  {year} {2008})}\BibitemShut {NoStop}%
\bibitem [{\citenamefont {Rotter}\ \emph {et~al.}(2008)\citenamefont {Rotter},
  \citenamefont {Tegel},\ and\ \citenamefont {Johrendt}}]{Rotter:2008a}%
  \BibitemOpen
  \bibfield  {author} {\bibinfo {author} {\bibfnamefont {M.}~\bibnamefont
  {Rotter}}, \bibinfo {author} {\bibfnamefont {M.}~\bibnamefont {Tegel}}, \
  and\ \bibinfo {author} {\bibfnamefont {D.}~\bibnamefont {Johrendt}},\ }\href
  {\doibase 10.1103/PhysRevLett.101.107006} {\bibfield  {journal} {\bibinfo
  {journal} {Phys. Rev. Lett.}\ }\textbf {\bibinfo {volume} {101}},\ \bibinfo
  {pages} {107006} (\bibinfo {year} {2008})}\BibitemShut {NoStop}%
\bibitem [{\citenamefont {Hirschfeld}\ \emph {et~al.}(2011)\citenamefont
  {Hirschfeld}, \citenamefont {Korshunov},\ and\ \citenamefont
  {Mazin}}]{Hirschfeld:2011}%
  \BibitemOpen
  \bibfield  {author} {\bibinfo {author} {\bibfnamefont {P.~J.}\ \bibnamefont
  {Hirschfeld}}, \bibinfo {author} {\bibfnamefont {M.~M.}\ \bibnamefont
  {Korshunov}}, \ and\ \bibinfo {author} {\bibfnamefont {I.~I.}\ \bibnamefont
  {Mazin}},\ }\href {\doibase doi:10.1088/0034-4885/74/12/124508} {\bibfield
  {journal} {\bibinfo  {journal} {Rep. Prog. Phys.}\ }\textbf {\bibinfo
  {volume} {74}},\ \bibinfo {pages} {124508} (\bibinfo {year}
  {2011})}\BibitemShut {NoStop}%
\bibitem [{\citenamefont {Lee}\ \emph {et~al.}(2009)\citenamefont {Lee},
  \citenamefont {Zhang},\ and\ \citenamefont {Wu}}]{Lee:2009}%
  \BibitemOpen
  \bibfield  {author} {\bibinfo {author} {\bibfnamefont {W.-C.}\ \bibnamefont
  {Lee}}, \bibinfo {author} {\bibfnamefont {S.-C.}\ \bibnamefont {Zhang}}, \
  and\ \bibinfo {author} {\bibfnamefont {C.}~\bibnamefont {Wu}},\ }\href
  {\doibase 10.1103/PhysRevLett.102.217002} {\bibfield  {journal} {\bibinfo
  {journal} {Phys. Rev. Lett.}\ }\textbf {\bibinfo {volume} {102}},\ \bibinfo
  {pages} {217002} (\bibinfo {year} {2009})}\BibitemShut {NoStop}%
\bibitem [{\citenamefont {Thomale}\ \emph
  {et~al.}(2011{\natexlab{a}})\citenamefont {Thomale}, \citenamefont {Platt},
  \citenamefont {Hanke},\ and\ \citenamefont {Bernevig}}]{Thomale:2011}%
  \BibitemOpen
  \bibfield  {author} {\bibinfo {author} {\bibfnamefont {R.}~\bibnamefont
  {Thomale}}, \bibinfo {author} {\bibfnamefont {C.}~\bibnamefont {Platt}},
  \bibinfo {author} {\bibfnamefont {W.}~\bibnamefont {Hanke}}, \ and\ \bibinfo
  {author} {\bibfnamefont {B.~A.}\ \bibnamefont {Bernevig}},\ }\href {\doibase
  10.1103/PhysRevLett.106.187003} {\bibfield  {journal} {\bibinfo  {journal}
  {Phys. Rev. Lett.}\ }\textbf {\bibinfo {volume} {106}},\ \bibinfo {pages}
  {187003} (\bibinfo {year} {2011}{\natexlab{a}})}\BibitemShut {NoStop}%
\bibitem [{\citenamefont {Mazin}\ \emph {et~al.}(2008)\citenamefont {Mazin},
  \citenamefont {Singh}, \citenamefont {Johannes},\ and\ \citenamefont
  {Du}}]{Mazin:2008a}%
  \BibitemOpen
  \bibfield  {author} {\bibinfo {author} {\bibfnamefont {I.~I.}\ \bibnamefont
  {Mazin}}, \bibinfo {author} {\bibfnamefont {D.~J.}\ \bibnamefont {Singh}},
  \bibinfo {author} {\bibfnamefont {M.~D.}\ \bibnamefont {Johannes}}, \ and\
  \bibinfo {author} {\bibfnamefont {M.~H.}\ \bibnamefont {Du}},\ }\href
  {\doibase 10.1103/PhysRevLett.101.057003} {\bibfield  {journal} {\bibinfo
  {journal} {Phys. Rev. Lett.}\ }\textbf {\bibinfo {volume} {101}},\ \bibinfo
  {eid} {057003} (\bibinfo {year} {2008})}\BibitemShut {NoStop}%
\bibitem [{\citenamefont {Kontani}\ and\ \citenamefont
  {Onari}(2010)}]{Kontani:2010}%
  \BibitemOpen
  \bibfield  {author} {\bibinfo {author} {\bibfnamefont {H.}~\bibnamefont
  {Kontani}}\ and\ \bibinfo {author} {\bibfnamefont {S.}~\bibnamefont
  {Onari}},\ }\href {\doibase 10.1103/PhysRevLett.104.157001} {\bibfield
  {journal} {\bibinfo  {journal} {Phys. Rev. Lett.}\ }\textbf {\bibinfo
  {volume} {104}},\ \bibinfo {pages} {157001} (\bibinfo {year}
  {2010})}\BibitemShut {NoStop}%
\bibitem [{\citenamefont {Scalapino}\ and\ \citenamefont
  {Devereaux}(2009)}]{Scalapino:2009}%
  \BibitemOpen
  \bibfield  {author} {\bibinfo {author} {\bibfnamefont {D.~J.}\ \bibnamefont
  {Scalapino}}\ and\ \bibinfo {author} {\bibfnamefont {T.~P.}\ \bibnamefont
  {Devereaux}},\ }\href {\doibase 10.1103/PhysRevB.80.140512} {\bibfield
  {journal} {\bibinfo  {journal} {Phys. Rev. B}\ }\textbf {\bibinfo {volume}
  {80}},\ \bibinfo {eid} {140512} (\bibinfo {year} {2009})}\BibitemShut
  {NoStop}%
\bibitem [{\citenamefont {Devereaux}(1994)}]{Devereaux:1994}%
  \BibitemOpen
  \bibfield  {author} {\bibinfo {author} {\bibfnamefont {T.~P.}\ \bibnamefont
  {Devereaux}},\ }\href {\doibase 10.1103/PhysRevB.50.10287} {\bibfield
  {journal} {\bibinfo  {journal} {Phys. Rev. B}\ }\textbf {\bibinfo {volume}
  {50}},\ \bibinfo {pages} {10287} (\bibinfo {year} {1994})}\BibitemShut
  {NoStop}%
\bibitem [{\citenamefont {Devereaux}\ and\ \citenamefont
  {Hackl}(2007)}]{Devereaux:2007}%
  \BibitemOpen
  \bibfield  {author} {\bibinfo {author} {\bibfnamefont {T.}~\bibnamefont
  {Devereaux}}\ and\ \bibinfo {author} {\bibfnamefont {R.}~\bibnamefont
  {Hackl}},\ }\href {\doibase 10.1103/RevModPhys.79.175} {\bibfield  {journal}
  {\bibinfo  {journal} {Rev. Mod. Phys.}\ }\textbf {\bibinfo {volume} {79}},\
  \bibinfo {pages} {175} (\bibinfo {year} {2007})}\BibitemShut {NoStop}%
\bibitem [{\citenamefont {Bardasis}\ and\ \citenamefont
  {Schrieffer}(1961)}]{Bardasis:1961}%
  \BibitemOpen
  \bibfield  {author} {\bibinfo {author} {\bibfnamefont {A.}~\bibnamefont
  {Bardasis}}\ and\ \bibinfo {author} {\bibfnamefont {J.~R.}\ \bibnamefont
  {Schrieffer}},\ }\href {\doibase 10.1103/PhysRev.121.1050} {\bibfield
  {journal} {\bibinfo  {journal} {Phys. Rev.}\ }\textbf {\bibinfo {volume}
  {121}},\ \bibinfo {pages} {1050} (\bibinfo {year} {1961})}\BibitemShut
  {NoStop}%
\bibitem [{\citenamefont {Klein}\ and\ \citenamefont
  {Dierker}(1984)}]{Klein:1984}%
  \BibitemOpen
  \bibfield  {author} {\bibinfo {author} {\bibfnamefont {M.~V.}\ \bibnamefont
  {Klein}}\ and\ \bibinfo {author} {\bibfnamefont {S.~B.}\ \bibnamefont
  {Dierker}},\ }\href {\doibase 10.1103/PhysRevB.29.4976} {\bibfield  {journal}
  {\bibinfo  {journal} {Phys. Rev. B}\ }\textbf {\bibinfo {volume} {29}},\
  \bibinfo {pages} {4976} (\bibinfo {year} {1984})}\BibitemShut {NoStop}%
\bibitem [{\citenamefont {Monien}\ and\ \citenamefont
  {Zawadowski}(1990)}]{Monien:1990}%
  \BibitemOpen
  \bibfield  {author} {\bibinfo {author} {\bibfnamefont {H.}~\bibnamefont
  {Monien}}\ and\ \bibinfo {author} {\bibfnamefont {A.}~\bibnamefont
  {Zawadowski}},\ }\href {\doibase 10.1103/PhysRevB.41.8798} {\bibfield
  {journal} {\bibinfo  {journal} {Phys. Rev. B}\ }\textbf {\bibinfo {volume}
  {41}},\ \bibinfo {pages} {8798} (\bibinfo {year} {1990})}\BibitemShut
  {NoStop}%
\bibitem [{\citenamefont {Chubukov}\ \emph {et~al.}(2009)\citenamefont
  {Chubukov}, \citenamefont {Eremin},\ and\ \citenamefont
  {Korshunov}}]{Chubukov:2009a}%
  \BibitemOpen
  \bibfield  {author} {\bibinfo {author} {\bibfnamefont {A.~V.}\ \bibnamefont
  {Chubukov}}, \bibinfo {author} {\bibfnamefont {I.}~\bibnamefont {Eremin}}, \
  and\ \bibinfo {author} {\bibfnamefont {M.~M.}\ \bibnamefont {Korshunov}},\
  }\href {\doibase 10.1103/PhysRevB.79.220501} {\bibfield  {journal} {\bibinfo
  {journal} {Phys. Rev. B}\ }\textbf {\bibinfo {volume} {79}},\ \bibinfo {eid}
  {220501} (\bibinfo {year} {2009})}\BibitemShut {NoStop}%
\bibitem [{\citenamefont {Ding}\ \emph {et~al.}(2008)\citenamefont {Ding},
  \citenamefont {Richard}, \citenamefont {Nakayama}, \citenamefont {Sugawara},
  \citenamefont {Arakane}, \citenamefont {Sekiba}, \citenamefont {Takayama},
  \citenamefont {Souma}, \citenamefont {Sato}, \citenamefont {Takahashi},
  \citenamefont {Wang}, \citenamefont {Dai}, \citenamefont {Fang},
  \citenamefont {Chen}, \citenamefont {Luo},\ and\ \citenamefont
  {Wang}}]{Ding:2008}%
  \BibitemOpen
  \bibfield  {author} {\bibinfo {author} {\bibfnamefont {H.}~\bibnamefont
  {Ding}}, \bibinfo {author} {\bibfnamefont {P.}~\bibnamefont {Richard}},
  \bibinfo {author} {\bibfnamefont {K.}~\bibnamefont {Nakayama}}, \bibinfo
  {author} {\bibfnamefont {K.}~\bibnamefont {Sugawara}}, \bibinfo {author}
  {\bibfnamefont {T.}~\bibnamefont {Arakane}}, \bibinfo {author} {\bibfnamefont
  {Y.}~\bibnamefont {Sekiba}}, \bibinfo {author} {\bibfnamefont
  {A.}~\bibnamefont {Takayama}}, \bibinfo {author} {\bibfnamefont
  {S.}~\bibnamefont {Souma}}, \bibinfo {author} {\bibfnamefont
  {T.}~\bibnamefont {Sato}}, \bibinfo {author} {\bibfnamefont {T.}~\bibnamefont
  {Takahashi}}, \bibinfo {author} {\bibfnamefont {Z.}~\bibnamefont {Wang}},
  \bibinfo {author} {\bibfnamefont {X.}~\bibnamefont {Dai}}, \bibinfo {author}
  {\bibfnamefont {Z.}~\bibnamefont {Fang}}, \bibinfo {author} {\bibfnamefont
  {G.~F.}\ \bibnamefont {Chen}}, \bibinfo {author} {\bibfnamefont {J.~L.}\
  \bibnamefont {Luo}}, \ and\ \bibinfo {author} {\bibfnamefont {N.~L.}\
  \bibnamefont {Wang}},\ }\href {\doibase 10.1209/0295-5075/83/47001}
  {\bibfield  {journal} {\bibinfo  {journal} {Europhys. Lett.}\ }\textbf
  {\bibinfo {volume} {83}},\ \bibinfo {pages} {47001} (\bibinfo {year}
  {2008})}\BibitemShut {NoStop}%
\bibitem [{\citenamefont {Evtushinsky}\ \emph {et~al.}(2009)\citenamefont
  {Evtushinsky}, \citenamefont {Inosov}, \citenamefont {Zabolotnyy},
  \citenamefont {Koitzsch}, \citenamefont {Knupfer}, \citenamefont {B\"uchner},
  \citenamefont {Viazovska}, \citenamefont {Sun}, \citenamefont {Hinkov},
  \citenamefont {Boris}, \citenamefont {Lin}, \citenamefont {Keimer},
  \citenamefont {Varykhalov}, \citenamefont {Kordyuk},\ and\ \citenamefont
  {Borisenko}}]{Evtushinsky:2009a}%
  \BibitemOpen
  \bibfield  {author} {\bibinfo {author} {\bibfnamefont {D.~V.}\ \bibnamefont
  {Evtushinsky}}, \bibinfo {author} {\bibfnamefont {D.~S.}\ \bibnamefont
  {Inosov}}, \bibinfo {author} {\bibfnamefont {V.~B.}\ \bibnamefont
  {Zabolotnyy}}, \bibinfo {author} {\bibfnamefont {A.}~\bibnamefont
  {Koitzsch}}, \bibinfo {author} {\bibfnamefont {M.}~\bibnamefont {Knupfer}},
  \bibinfo {author} {\bibfnamefont {B.}~\bibnamefont {B\"uchner}}, \bibinfo
  {author} {\bibfnamefont {M.~S.}\ \bibnamefont {Viazovska}}, \bibinfo {author}
  {\bibfnamefont {G.~L.}\ \bibnamefont {Sun}}, \bibinfo {author} {\bibfnamefont
  {V.}~\bibnamefont {Hinkov}}, \bibinfo {author} {\bibfnamefont {A.~V.}\
  \bibnamefont {Boris}}, \bibinfo {author} {\bibfnamefont {C.~T.}\ \bibnamefont
  {Lin}}, \bibinfo {author} {\bibfnamefont {B.}~\bibnamefont {Keimer}},
  \bibinfo {author} {\bibfnamefont {A.}~\bibnamefont {Varykhalov}}, \bibinfo
  {author} {\bibfnamefont {A.~A.}\ \bibnamefont {Kordyuk}}, \ and\ \bibinfo
  {author} {\bibfnamefont {S.~V.}\ \bibnamefont {Borisenko}},\ }\href {\doibase
  10.1103/PhysRevB.79.054517} {\bibfield  {journal} {\bibinfo  {journal} {Phys.
  Rev. B}\ }\textbf {\bibinfo {volume} {79}},\ \bibinfo {eid} {054517}
  (\bibinfo {year} {2009})}\BibitemShut {NoStop}%
\bibitem [{\citenamefont {Zhang}\ \emph {et~al.}(2011)\citenamefont {Zhang},
  \citenamefont {Yang}, \citenamefont {Xu}, \citenamefont {Ye}, \citenamefont
  {Chen}, \citenamefont {He}, \citenamefont {Xu}, \citenamefont {Jiang},
  \citenamefont {Xie}, \citenamefont {Ying}, \citenamefont {Wang},
  \citenamefont {Chen}, \citenamefont {Hu}, \citenamefont {Matsunami},
  \citenamefont {Kimura},\ and\ \citenamefont {Feng}}]{Zhang:2011a}%
  \BibitemOpen
  \bibfield  {author} {\bibinfo {author} {\bibfnamefont {Y.}~\bibnamefont
  {Zhang}}, \bibinfo {author} {\bibfnamefont {L.~X.}\ \bibnamefont {Yang}},
  \bibinfo {author} {\bibfnamefont {M.}~\bibnamefont {Xu}}, \bibinfo {author}
  {\bibfnamefont {Z.~R.}\ \bibnamefont {Ye}}, \bibinfo {author} {\bibfnamefont
  {F.}~\bibnamefont {Chen}}, \bibinfo {author} {\bibfnamefont {C.}~\bibnamefont
  {He}}, \bibinfo {author} {\bibfnamefont {H.~C.}\ \bibnamefont {Xu}}, \bibinfo
  {author} {\bibfnamefont {J.}~\bibnamefont {Jiang}}, \bibinfo {author}
  {\bibfnamefont {B.~P.}\ \bibnamefont {Xie}}, \bibinfo {author} {\bibfnamefont
  {J.~J.}\ \bibnamefont {Ying}}, \bibinfo {author} {\bibfnamefont {X.~F.}\
  \bibnamefont {Wang}}, \bibinfo {author} {\bibfnamefont {X.~H.}\ \bibnamefont
  {Chen}}, \bibinfo {author} {\bibfnamefont {J.~P.}\ \bibnamefont {Hu}},
  \bibinfo {author} {\bibfnamefont {M.}~\bibnamefont {Matsunami}}, \bibinfo
  {author} {\bibfnamefont {S.}~\bibnamefont {Kimura}}, \ and\ \bibinfo {author}
  {\bibfnamefont {D.~L.}\ \bibnamefont {Feng}},\ }\href
  {http://dx.doi.org/10.1038/nmat2981} {\bibfield  {journal} {\bibinfo
  {journal} {Nat. Mater.}\ }\textbf {\bibinfo {volume} {10}},\ \bibinfo {pages}
  {273} (\bibinfo {year} {2011})}\BibitemShut {NoStop}%
\bibitem [{\citenamefont {Qian}\ \emph {et~al.}(2011)\citenamefont {Qian},
  \citenamefont {Wang}, \citenamefont {Jin}, \citenamefont {Zhang},
  \citenamefont {Richard}, \citenamefont {Xu}, \citenamefont {Dai},
  \citenamefont {Fang}, \citenamefont {Guo}, \citenamefont {Chen},\ and\
  \citenamefont {Ding}}]{Qian:2011}%
  \BibitemOpen
  \bibfield  {author} {\bibinfo {author} {\bibfnamefont {T.}~\bibnamefont
  {Qian}}, \bibinfo {author} {\bibfnamefont {X.-P.}\ \bibnamefont {Wang}},
  \bibinfo {author} {\bibfnamefont {W.-C.}\ \bibnamefont {Jin}}, \bibinfo
  {author} {\bibfnamefont {P.}~\bibnamefont {Zhang}}, \bibinfo {author}
  {\bibfnamefont {P.}~\bibnamefont {Richard}}, \bibinfo {author} {\bibfnamefont
  {G.}~\bibnamefont {Xu}}, \bibinfo {author} {\bibfnamefont {X.}~\bibnamefont
  {Dai}}, \bibinfo {author} {\bibfnamefont {Z.}~\bibnamefont {Fang}}, \bibinfo
  {author} {\bibfnamefont {J.-G.}\ \bibnamefont {Guo}}, \bibinfo {author}
  {\bibfnamefont {X.-L.}\ \bibnamefont {Chen}}, \ and\ \bibinfo {author}
  {\bibfnamefont {H.}~\bibnamefont {Ding}},\ }\href {\doibase
  10.1103/PhysRevLett.106.187001} {\bibfield  {journal} {\bibinfo  {journal}
  {Phys. Rev. Lett.}\ }\textbf {\bibinfo {volume} {106}},\ \bibinfo {pages}
  {187001} (\bibinfo {year} {2011})}\BibitemShut {NoStop}%
\bibitem [{\citenamefont {Muschler}\ \emph {et~al.}(2009)\citenamefont
  {Muschler}, \citenamefont {Prestel}, \citenamefont {Hackl}, \citenamefont
  {Devereaux}, \citenamefont {Analytis}, \citenamefont {Chu},\ and\
  \citenamefont {Fisher}}]{Muschler:2009}%
  \BibitemOpen
  \bibfield  {author} {\bibinfo {author} {\bibfnamefont {B.}~\bibnamefont
  {Muschler}}, \bibinfo {author} {\bibfnamefont {W.}~\bibnamefont {Prestel}},
  \bibinfo {author} {\bibfnamefont {R.}~\bibnamefont {Hackl}}, \bibinfo
  {author} {\bibfnamefont {T.~P.}\ \bibnamefont {Devereaux}}, \bibinfo {author}
  {\bibfnamefont {J.~G.}\ \bibnamefont {Analytis}}, \bibinfo {author}
  {\bibfnamefont {J.-H.}\ \bibnamefont {Chu}}, \ and\ \bibinfo {author}
  {\bibfnamefont {I.~R.}\ \bibnamefont {Fisher}},\ }\href {\doibase
  10.1103/PhysRevB.80.180510} {\bibfield  {journal} {\bibinfo  {journal} {Phys.
  Rev. B}\ }\textbf {\bibinfo {volume} {80}},\ \bibinfo {pages} {180510}
  (\bibinfo {year} {2009})}\BibitemShut {NoStop}%
\bibitem [{\citenamefont {Shen}\ \emph {et~al.}(2011)\citenamefont {Shen},
  \citenamefont {Yang}, \citenamefont {Wang}, \citenamefont {Han},
  \citenamefont {Zeng}, \citenamefont {Shan}, \citenamefont {Ren},\ and\
  \citenamefont {Wen}}]{Shen:2011}%
  \BibitemOpen
  \bibfield  {author} {\bibinfo {author} {\bibfnamefont {B.}~\bibnamefont
  {Shen}}, \bibinfo {author} {\bibfnamefont {H.}~\bibnamefont {Yang}}, \bibinfo
  {author} {\bibfnamefont {Z.-S.}\ \bibnamefont {Wang}}, \bibinfo {author}
  {\bibfnamefont {F.}~\bibnamefont {Han}}, \bibinfo {author} {\bibfnamefont
  {B.}~\bibnamefont {Zeng}}, \bibinfo {author} {\bibfnamefont {L.}~\bibnamefont
  {Shan}}, \bibinfo {author} {\bibfnamefont {C.}~\bibnamefont {Ren}}, \ and\
  \bibinfo {author} {\bibfnamefont {H.-H.}\ \bibnamefont {Wen}},\ }\href
  {\doibase 10.1103/PhysRevB.84.184512} {\bibfield  {journal} {\bibinfo
  {journal} {Phys. Rev. B}\ }\textbf {\bibinfo {volume} {84}},\ \bibinfo
  {pages} {184512} (\bibinfo {year} {2011})}\BibitemShut {NoStop}%
\bibitem [{\citenamefont {Tsurkan}\ \emph {et~al.}(2011)\citenamefont
  {Tsurkan}, \citenamefont {Deisenhofer}, \citenamefont {G\"unther},
  \citenamefont {Krug~von Nidda}, \citenamefont {Widmann},\ and\ \citenamefont
  {Loidl}}]{Tsurkan:2011}%
  \BibitemOpen
  \bibfield  {author} {\bibinfo {author} {\bibfnamefont {V.}~\bibnamefont
  {Tsurkan}}, \bibinfo {author} {\bibfnamefont {J.}~\bibnamefont
  {Deisenhofer}}, \bibinfo {author} {\bibfnamefont {A.}~\bibnamefont
  {G\"unther}}, \bibinfo {author} {\bibfnamefont {H.-A.}\ \bibnamefont
  {Krug~von Nidda}}, \bibinfo {author} {\bibfnamefont {S.}~\bibnamefont
  {Widmann}}, \ and\ \bibinfo {author} {\bibfnamefont {A.}~\bibnamefont
  {Loidl}},\ }\href {\doibase 10.1103/PhysRevB.84.144520} {\bibfield  {journal}
  {\bibinfo  {journal} {Phys. Rev. B}\ }\textbf {\bibinfo {volume} {84}},\
  \bibinfo {pages} {144520} (\bibinfo {year} {2011})}\BibitemShut {NoStop}%
\bibitem [{\citenamefont {Ksenofontov}\ \emph {et~al.}(2011)\citenamefont
  {Ksenofontov}, \citenamefont {Wortmann}, \citenamefont {Medvedev},
  \citenamefont {Tsurkan}, \citenamefont {Deisenhofer}, \citenamefont {Loidl},\
  and\ \citenamefont {Felser}}]{Ksenofontov:2011}%
  \BibitemOpen
  \bibfield  {author} {\bibinfo {author} {\bibfnamefont {V.}~\bibnamefont
  {Ksenofontov}}, \bibinfo {author} {\bibfnamefont {G.}~\bibnamefont
  {Wortmann}}, \bibinfo {author} {\bibfnamefont {S.~A.}\ \bibnamefont
  {Medvedev}}, \bibinfo {author} {\bibfnamefont {V.}~\bibnamefont {Tsurkan}},
  \bibinfo {author} {\bibfnamefont {J.}~\bibnamefont {Deisenhofer}}, \bibinfo
  {author} {\bibfnamefont {A.}~\bibnamefont {Loidl}}, \ and\ \bibinfo {author}
  {\bibfnamefont {C.}~\bibnamefont {Felser}},\ }\href {\doibase
  10.1103/PhysRevB.84.180508} {\bibfield  {journal} {\bibinfo  {journal} {Phys.
  Rev. B}\ }\textbf {\bibinfo {volume} {84}},\ \bibinfo {pages} {180508}
  (\bibinfo {year} {2011})}\BibitemShut {NoStop}%
\bibitem [{\citenamefont {Chauvi\`ere}\ \emph {et~al.}(2010)\citenamefont
  {Chauvi\`ere}, \citenamefont {Gallais}, \citenamefont {Cazayous},
  \citenamefont {M\'easson}, \citenamefont {Sacuto}, \citenamefont {Colson},\
  and\ \citenamefont {Forget}}]{Chauviere:2010}%
  \BibitemOpen
  \bibfield  {author} {\bibinfo {author} {\bibfnamefont {L.}~\bibnamefont
  {Chauvi\`ere}}, \bibinfo {author} {\bibfnamefont {Y.}~\bibnamefont
  {Gallais}}, \bibinfo {author} {\bibfnamefont {M.}~\bibnamefont {Cazayous}},
  \bibinfo {author} {\bibfnamefont {M.~A.}\ \bibnamefont {M\'easson}}, \bibinfo
  {author} {\bibfnamefont {A.}~\bibnamefont {Sacuto}}, \bibinfo {author}
  {\bibfnamefont {D.}~\bibnamefont {Colson}}, \ and\ \bibinfo {author}
  {\bibfnamefont {A.}~\bibnamefont {Forget}},\ }\href {\doibase
  10.1103/PhysRevB.82.180521} {\bibfield  {journal} {\bibinfo  {journal} {Phys.
  Rev. B}\ }\textbf {\bibinfo {volume} {82}},\ \bibinfo {pages} {180521}
  (\bibinfo {year} {2010})}\BibitemShut {NoStop}%
\bibitem [{\citenamefont {Mazin}\ \emph {et~al.}(2010)\citenamefont {Mazin},
  \citenamefont {Devereaux}, \citenamefont {Analytis}, \citenamefont {Chu},
  \citenamefont {Fisher}, \citenamefont {Muschler},\ and\ \citenamefont
  {Hackl}}]{Mazin:2010a}%
  \BibitemOpen
  \bibfield  {author} {\bibinfo {author} {\bibfnamefont {I.~I.}\ \bibnamefont
  {Mazin}}, \bibinfo {author} {\bibfnamefont {T.~P.}\ \bibnamefont
  {Devereaux}}, \bibinfo {author} {\bibfnamefont {J.~G.}\ \bibnamefont
  {Analytis}}, \bibinfo {author} {\bibfnamefont {J.-H.}\ \bibnamefont {Chu}},
  \bibinfo {author} {\bibfnamefont {I.~R.}\ \bibnamefont {Fisher}}, \bibinfo
  {author} {\bibfnamefont {B.}~\bibnamefont {Muschler}}, \ and\ \bibinfo
  {author} {\bibfnamefont {R.}~\bibnamefont {Hackl}},\ }\href {\doibase
  10.1103/PhysRevB.82.180502} {\bibfield  {journal} {\bibinfo  {journal} {Phys.
  Rev. B}\ }\textbf {\bibinfo {volume} {82}},\ \bibinfo {pages} {180502}
  (\bibinfo {year} {2010})}\BibitemShut {NoStop}%
\bibitem [{\citenamefont {Platt}\ \emph {et~al.}(2012)\citenamefont {Platt},
  \citenamefont {Thomale}, \citenamefont {Honerkamp}, \citenamefont {Zhang},\
  and\ \citenamefont {Hanke}}]{Platt:2012}%
  \BibitemOpen
  \bibfield  {author} {\bibinfo {author} {\bibfnamefont {C.}~\bibnamefont
  {Platt}}, \bibinfo {author} {\bibfnamefont {R.}~\bibnamefont {Thomale}},
  \bibinfo {author} {\bibfnamefont {C.}~\bibnamefont {Honerkamp}}, \bibinfo
  {author} {\bibfnamefont {S.-C.}\ \bibnamefont {Zhang}}, \ and\ \bibinfo
  {author} {\bibfnamefont {W.}~\bibnamefont {Hanke}},\ }\href {\doibase
  10.1103/PhysRevB.85.180502} {\bibfield  {journal} {\bibinfo  {journal} {Phys.
  Rev. B}\ }\textbf {\bibinfo {volume} {85}},\ \bibinfo {pages} {180502}
  (\bibinfo {year} {2012})}\BibitemShut {NoStop}%
\bibitem [{\citenamefont {Kuroki}\ \emph {et~al.}(2009)\citenamefont {Kuroki},
  \citenamefont {Usui}, \citenamefont {Onari}, \citenamefont {Arita},\ and\
  \citenamefont {Aoki}}]{Kuroki:2009}%
  \BibitemOpen
  \bibfield  {author} {\bibinfo {author} {\bibfnamefont {K.}~\bibnamefont
  {Kuroki}}, \bibinfo {author} {\bibfnamefont {H.}~\bibnamefont {Usui}},
  \bibinfo {author} {\bibfnamefont {S.}~\bibnamefont {Onari}}, \bibinfo
  {author} {\bibfnamefont {R.}~\bibnamefont {Arita}}, \ and\ \bibinfo {author}
  {\bibfnamefont {H.}~\bibnamefont {Aoki}},\ }\href {\doibase
  10.1103/PhysRevB.79.224511} {\bibfield  {journal} {\bibinfo  {journal} {Phys.
  Rev. B}\ }\textbf {\bibinfo {volume} {79}},\ \bibinfo {pages} {224511}
  (\bibinfo {year} {2009})}\BibitemShut {NoStop}%
\bibitem [{\citenamefont {Ikeda}\ \emph {et~al.}(2010)\citenamefont {Ikeda},
  \citenamefont {Arita},\ and\ \citenamefont {Kune\ifmmode~\check{s}\else
  \v{s}\fi{}}}]{Ikeda:2010}%
  \BibitemOpen
  \bibfield  {author} {\bibinfo {author} {\bibfnamefont {H.}~\bibnamefont
  {Ikeda}}, \bibinfo {author} {\bibfnamefont {R.}~\bibnamefont {Arita}}, \ and\
  \bibinfo {author} {\bibfnamefont {J.}~\bibnamefont
  {Kune\ifmmode~\check{s}\else \v{s}\fi{}}},\ }\href {\doibase
  10.1103/PhysRevB.81.054502} {\bibfield  {journal} {\bibinfo  {journal} {Phys.
  Rev. B}\ }\textbf {\bibinfo {volume} {81}},\ \bibinfo {pages} {054502}
  (\bibinfo {year} {2010})}\BibitemShut {NoStop}%
\bibitem [{\citenamefont {Maiti}\ \emph {et~al.}(2011)\citenamefont {Maiti},
  \citenamefont {Korshunov}, \citenamefont {Maier}, \citenamefont
  {Hirschfeld},\ and\ \citenamefont {Chubukov}}]{Maiti:2011a}%
  \BibitemOpen
  \bibfield  {author} {\bibinfo {author} {\bibfnamefont {S.}~\bibnamefont
  {Maiti}}, \bibinfo {author} {\bibfnamefont {M.~M.}\ \bibnamefont
  {Korshunov}}, \bibinfo {author} {\bibfnamefont {T.~A.}\ \bibnamefont
  {Maier}}, \bibinfo {author} {\bibfnamefont {P.~J.}\ \bibnamefont
  {Hirschfeld}}, \ and\ \bibinfo {author} {\bibfnamefont {A.~V.}\ \bibnamefont
  {Chubukov}},\ }\href {\doibase 10.1103/PhysRevLett.107.147002} {\bibfield
  {journal} {\bibinfo  {journal} {Phys. Rev. Lett.}\ }\textbf {\bibinfo
  {volume} {107}},\ \bibinfo {pages} {147002} (\bibinfo {year}
  {2011})}\BibitemShut {NoStop}%
\bibitem [{\citenamefont {{Das}}\ and\ \citenamefont
  {{Balatsky}}(2012)}]{Das:2012}%
  \BibitemOpen
  \bibfield  {author} {\bibinfo {author} {\bibfnamefont {T.}~\bibnamefont
  {{Das}}}\ and\ \bibinfo {author} {\bibfnamefont {A.~V.}\ \bibnamefont
  {{Balatsky}}},\ }\href@noop {} {\bibfield  {journal} {\bibinfo  {journal}
  {ArXiv e-prints}\ } (\bibinfo {year} {2012})},\ \Eprint
  {http://arxiv.org/abs/1208.2468} {arXiv:1208.2468 [cond-mat.supr-con]}
  \BibitemShut {NoStop}%
\bibitem [{\citenamefont {{Fernandes}}\ and\ \citenamefont
  {{Millis}}(2012)}]{Fernandes:2012a}%
  \BibitemOpen
  \bibfield  {author} {\bibinfo {author} {\bibfnamefont {R.~M.}\ \bibnamefont
  {{Fernandes}}}\ and\ \bibinfo {author} {\bibfnamefont {A.~J.}\ \bibnamefont
  {{Millis}}},\ }\href@noop {} {\bibfield  {journal} {\bibinfo  {journal}
  {ArXiv e-prints}\ } (\bibinfo {year} {2012})},\ \Eprint
  {http://arxiv.org/abs/1208.3412} {arXiv:1208.3412 [cond-mat.supr-con]}
  \BibitemShut {NoStop}%
\bibitem [{\citenamefont {Thomale}\ \emph
  {et~al.}(2011{\natexlab{b}})\citenamefont {Thomale}, \citenamefont {Platt},
  \citenamefont {Hanke}, \citenamefont {Hu},\ and\ \citenamefont
  {Bernevig}}]{Thomale:2011a}%
  \BibitemOpen
  \bibfield  {author} {\bibinfo {author} {\bibfnamefont {R.}~\bibnamefont
  {Thomale}}, \bibinfo {author} {\bibfnamefont {C.}~\bibnamefont {Platt}},
  \bibinfo {author} {\bibfnamefont {W.}~\bibnamefont {Hanke}}, \bibinfo
  {author} {\bibfnamefont {J.}~\bibnamefont {Hu}}, \ and\ \bibinfo {author}
  {\bibfnamefont {B.~A.}\ \bibnamefont {Bernevig}},\ }\href {\doibase
  10.1103/PhysRevLett.107.117001} {\bibfield  {journal} {\bibinfo  {journal}
  {Phys. Rev. Lett.}\ }\textbf {\bibinfo {volume} {107}},\ \bibinfo {pages}
  {117001} (\bibinfo {year} {2011}{\natexlab{b}})}\BibitemShut {NoStop}%
\bibitem [{\citenamefont {Abrikosov}\ and\ \citenamefont
  {Fal'kovskii}(1961)}]{Abrikosov:1961}%
  \BibitemOpen
  \bibfield  {author} {\bibinfo {author} {\bibfnamefont {A.~A.}\ \bibnamefont
  {Abrikosov}}\ and\ \bibinfo {author} {\bibfnamefont {L.~A.}\ \bibnamefont
  {Fal'kovskii}},\ }\href@noop {} {\bibfield  {journal} {\bibinfo  {journal}
  {Zh. Eksp. Teor. Fiz.}\ }\textbf {\bibinfo {volume} {40}},\ \bibinfo {pages}
  {262} (\bibinfo {year} {1961})},\ \bibinfo {note} {[Sov. Phys. JETP 13, 179
  (1961)]}\BibitemShut {NoStop}%
\bibitem [{\citenamefont {Devereaux}\ and\ \citenamefont
  {Einzel}(1995)}]{Devereaux:1995a}%
  \BibitemOpen
  \bibfield  {author} {\bibinfo {author} {\bibfnamefont {T.~P.}\ \bibnamefont
  {Devereaux}}\ and\ \bibinfo {author} {\bibfnamefont {D.}~\bibnamefont
  {Einzel}},\ }\href {\doibase 10.1103/PhysRevB.51.16336} {\bibfield  {journal}
  {\bibinfo  {journal} {Phys. Rev. B}\ }\textbf {\bibinfo {volume} {51}},\
  \bibinfo {pages} {16336} (\bibinfo {year} {1995})}\BibitemShut {NoStop}%
\bibitem [{\citenamefont {Mazin}(2011)}]{Mazin:2011a}%
  \BibitemOpen
  \bibfield  {author} {\bibinfo {author} {\bibfnamefont {I.~I.}\ \bibnamefont
  {Mazin}},\ }\href {\doibase 10.1103/PhysRevB.84.024529} {\bibfield  {journal}
  {\bibinfo  {journal} {Phys. Rev. B}\ }\textbf {\bibinfo {volume} {84}},\
  \bibinfo {pages} {024529} (\bibinfo {year} {2011})}\BibitemShut {NoStop}%
\bibitem [{\citenamefont {Khodas}\ and\ \citenamefont
  {Chubukov}(2012)}]{Khodas:2012}%
  \BibitemOpen
  \bibfield  {author} {\bibinfo {author} {\bibfnamefont {M.}~\bibnamefont
  {Khodas}}\ and\ \bibinfo {author} {\bibfnamefont {A.~V.}\ \bibnamefont
  {Chubukov}},\ }\href {\doibase 10.1103/PhysRevLett.108.247003} {\bibfield
  {journal} {\bibinfo  {journal} {Phys. Rev. Lett.}\ }\textbf {\bibinfo
  {volume} {108}},\ \bibinfo {pages} {247003} (\bibinfo {year}
  {2012})}\BibitemShut {NoStop}%
\bibitem [{\citenamefont {{Barlas}}\ and\ \citenamefont
  {{Varma}}(2012)}]{Barlas:2012}%
  \BibitemOpen
  \bibfield  {author} {\bibinfo {author} {\bibfnamefont {Y.}~\bibnamefont
  {{Barlas}}}\ and\ \bibinfo {author} {\bibfnamefont {C.~M.}\ \bibnamefont
  {{Varma}}},\ }\href@noop {} {\bibfield  {journal} {\bibinfo  {journal} {ArXiv
  e-prints}\ } (\bibinfo {year} {2012})},\ \Eprint
  {http://arxiv.org/abs/1206.0400} {arXiv:1206.0400 [cond-mat.supr-con]}
  \BibitemShut {NoStop}%
\end{thebibliography}
%\bibliographystyle{apsrev4-1}


\end{document}